\documentclass[a4paper,fleqn]{cas-sc}

\usepackage[authoryear,longnamesfirst]{natbib}

\def\tsc#1{\csdef{#1}{\textsc{\lowercase{#1}}\xspace}}
\tsc{WGM}
\tsc{QE}
\usepackage{lineno}
\usepackage{graphicx}
\usepackage{multirow}
\usepackage{amsmath,amssymb,amsfonts}%

\usepackage{algorithm}
\usepackage{algorithmicx}
\usepackage{algpseudocode}
\usepackage{float}
\usepackage{amsthm}
\usepackage{mathrsfs}
\usepackage[title]{appendix}
\usepackage{xcolor}
\usepackage{textcomp}
\usepackage{manyfoot}
\usepackage{booktabs}
\usepackage{algorithm}
\usepackage{algorithmicx}
\usepackage{algpseudocode}
\usepackage{listings}
\usepackage{array}
\usepackage{siunitx}
\usepackage{threeparttable} 
\usepackage{subcaption}
\usepackage{rotating}
\usepackage{hyperref}
\usepackage{tcolorbox}
\usepackage{tabularx}

\begin{document}
\let\WriteBookmarks\relax
\def\floatpagepagefraction{1}
\def\textpagefraction{.001}

\shorttitle{Causal Digital Twins for Cyber-Physical Security in Water Systems}    

\shortauthors{Homaei, et al.}  

\title [mode = title]{Causal Digital Twins for Cyber-Physical Security in Water Systems: A Framework for Robust Anomaly Detection}




\author[1]{Mohammadhossein Homaei}
\cormark[1] % Marca de autor correspondiente
\ead{mhomaein@alumnos.unex.es}

% Autor 2
\author[2]{Mehran Tarif}
\ead{mehran.tarifhokmabadi@univr.it}

% Autor 3
\author[1]{Pablo Garc\'{i}a Rodr\'{i}guez}
\ead{pablogr@unex.es}

% Autor 4
\author[1]{Andr\'{e}s Caro}
\ead{andresc@unex.es}

% Autor 5
\author[1]{Mar \'{A}vila}
\ead{mmavila@unex.es}

% Afiliación 1 (unex)
\affiliation[1]{organization={Department of Computer Systems and Telematics Engineering, University of Extremadura},
             addressline={Av. Universidad s/n}, 
             city={C\'{a}ceres},
             postcode={10003}, 
             country={Spain}}

% Afiliación 2 (univr)
\affiliation[2]{organization={Department of Computer Science, University of Verona},
             addressline={Strada le Grazie 15}, % Dirección más específica
             city={Verona},
             postcode={37134}, % Código postal correcto de Verona
             country={Italy}}

% Texto del autor correspondiente
\cortext[1]{Corresponding author.}



\begin{abstract}
Industrial Control Systems (ICS) in water distribution and treatment face cyber-physical attacks exploiting network and physical vulnerabilities. Current water system anomaly detection methods rely on correlations, yielding high false alarms and poor root cause analysis. We propose a Causal Digital Twin (CDT) framework for water infrastructures, combining causal inference with digital twin modeling. CDT supports association for pattern detection, intervention for system response, and counterfactual analysis for water attack prevention. Evaluated on water-related datasets SWaT, WADI, and HAI, CDT shows 90.8\% compliance with physical constraints and structural Hamming distance 0.133 $\pm$ 0.02. F1-scores are $0.944 \pm 0.014$ (SWaT), $0.902 \pm 0.021$ (WADI), $0.923 \pm 0.018$ (HAI, $p<0.0024$). CDT reduces false positives by 74\%, achieves 78.4\% root cause accuracy, and enables counterfactual defenses reducing attack success by 73.2\%. Real-time performance at 3.2 ms latency ensures safe and interpretable operation for medium-scale water systems.
\end{abstract}

\begin{graphicalabstract}
    \begin{tcolorbox}[
        colback=white, 
        colframe=black, 
        width=\linewidth, 
        boxrule=0.8pt, 
        arc=2mm, 
        title=Graphical Overview, 
        fonttitle=\bfseries
    ]
        \centering
        \includegraphics[width=0.9\linewidth]{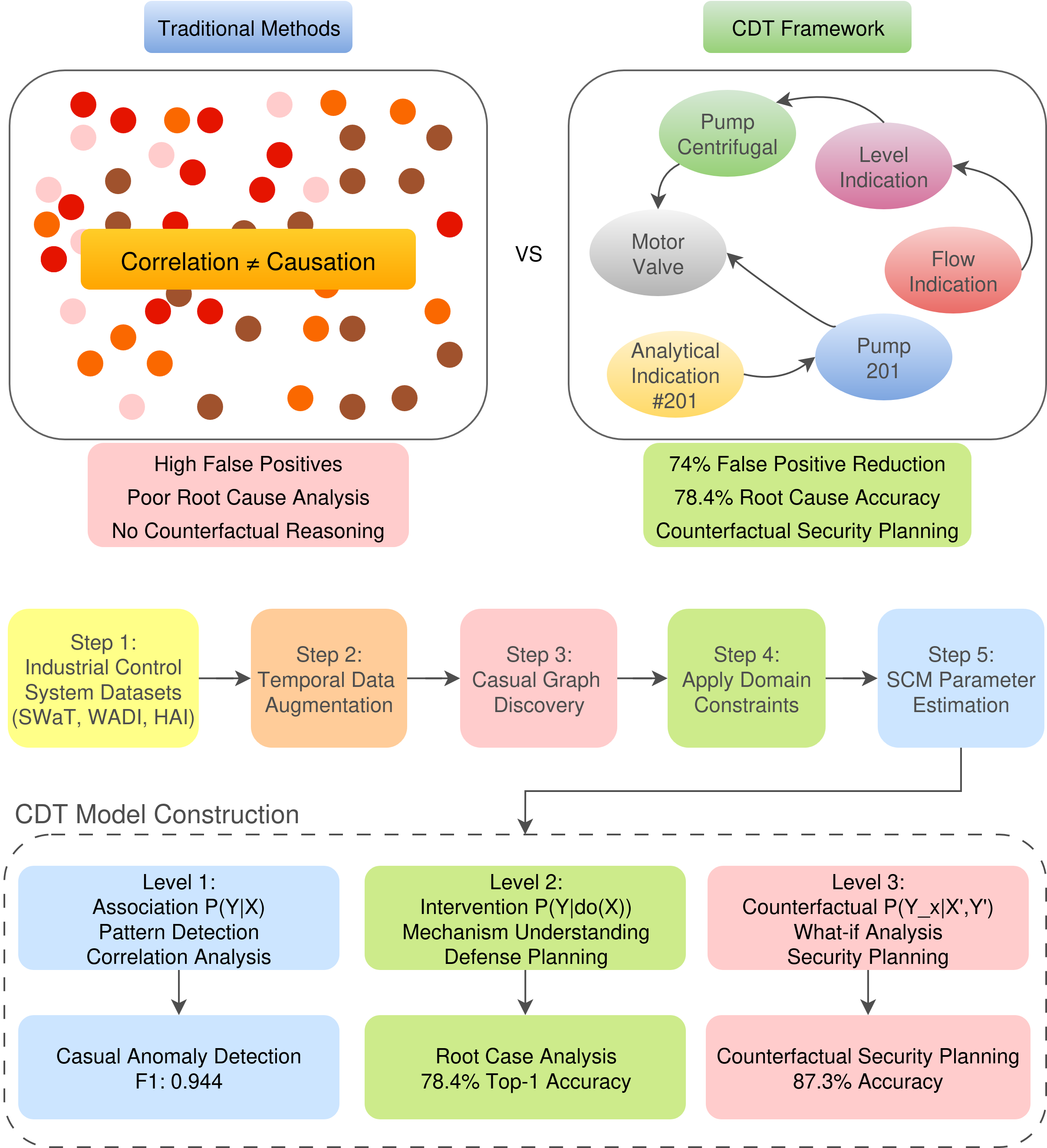}
    \end{tcolorbox}
\end{graphicalabstract}

\begin{highlights}
\item Novel CDT framework combines causal inference theory with digital twin modeling to enable associational, interventional, and counterfactual reasoning for cyber-physical security in ICS
\item Achieves statistically significant performance improvements with F1-scores of 0.944±0.014 (SWaT), 0.902±0.021 (WADI), and 0.923±0.018 (HAI), while reducing false positives by 74\% compared to correlation-based methods
\item Automated causal structure discovery with 90.8\% physical constraint compliance enables accurate root cause analysis achieving 78.4\% Top-1 accuracy, improving 29.7\% over traditional attribution methods
\item Counterfactual analysis capabilities provide proactive defense planning, identifying strategies that reduce attack success rates by 73.2\% across 84 documented attack scenarios
\item Real-time deployment feasibility with 3.2ms inference latency and successful cross-dataset transfer (F1-scores 0.847-0.889), demonstrating industrial scalability for medium-scale control systems.
\end{highlights}

% Keywords

\maketitle

\section{Introduction}\label{sec1:introduction}
The ICS have changed from isolated proprietary networks to interconnected cyber-physical systems that combine operational technology with information technology infrastructure. This change started in the 1990s with the use of standard networking protocols and has grown faster with Industry 4.0 initiatives. Today, ICS need Internet connectivity for remote monitoring, predictive maintenance, and distributed control across facilities that are far apart~\cite{Mathur2016,Xu2023,Abdoli2023,Jiang2024}. Critical infrastructure sectors such as water treatment, power generation, and manufacturing depend more on these connected systems, with the global ICS market expected to reach \$124 billion by 2025. The Secure Water Treatment (SWaT) testbed, developed by iTrust at Singapore University of Technology and Design, is an important example for cyber-physical security research. It provides a realistic scaled-down industrial environment with 51 sensors and actuators across six operational stages~\cite{Goh2017,Jeremiah2024,Lucchese2024}.

However, more connectivity and complexity create new cybersecurity vulnerabilities that traditional IT security cannot fully solve. The 2010 Stuxnet attack on Iranian nuclear facilities showed that advanced attackers can exploit both cyber weaknesses and physical process knowledge to cause real damage~\cite{Raman2022}. Recent events, like the 2021 Colonial Pipeline ransomware attack and the 2021 Oldsmar water treatment plant breach, show the ongoing threat to critical infrastructure. Current anomaly detection methods in ICS mostly use statistical correlation and machine learning techniques, which cannot separate real causal relationships from false associations~\cite{Kravchik2018,Kim2023}. These correlation-based methods often have high false positive rates, cannot do root cause analysis, and are vulnerable to adversarial attacks that take advantage of missing causal understanding.

The main limitation of current methods is their inability to reason about causal relationships in cyber-physical systems. Correlation can show statistical dependencies between sensor readings, but it cannot tell if anomalies come from attacks, cascading failures, or normal operational changes. This problem is serious when security analysts need actionable information for incident response, because correlation-based alerts do not provide enough guidance to prevent damage or recover the system. Also, without counterfactual reasoning, it is not possible to plan proactively or perform what-if analyses that could strengthen security before attacks happen.

In this research, we develop a comprehensive CDT framework for cyber-physical security in medium-scale ICS. Our method combines causal inference theory with DT modeling to allow three types of causal reasoning: associational queries for pattern recognition, interventional queries to understand system reactions to defensive actions, and counterfactual queries for analyzing attacks retrospectively and planning prevention. We test our framework using the SWaT dataset, showing better performance than correlation-based methods over 41 cyber-physical attack scenarios, while also giving interpretable root cause analysis and actionable security recommendations.

The rest of this paper is organized as following. Section~\ref{sec2:related_works} gives a review of existing methods for ICS security and DT applications, and it shows the limitations of correlation-based methods. Section~\ref{sec3:problem} defines the research problem and introduces the basic ideas of causal reasoning for cyber-physical security. Section~\ref{sec4:methodology} describes our newCDT framework, including causal graph discovery, estimation of structural causal model, and interventional reasoning algorithms. Section~\ref{sec5:results} shows the evaluation of the framework on the SWaT testbed compared with state-of-the-art methods. Finally, Section~\ref{sec6:conclusion} finishes the paper with a summary of main contributions, discussion of practical implications, limitations of the work, and suggestions for future research.

\section{Related Works}\label{sec2:related_works}

\subsection{DTs in Critical Infrastructure and ICS Security}

DTs have become an important tool for modeling and monitoring critical infrastructure systems, giving real-time virtual representations of physical assets and processes \cite{Homaei2022, Homaei2024review,Homaei2024Digital}. Xu et al.~\cite{Xu2023} made a detailed survey of DT applications in Industrial Internet of Things (IoT). They identified key technologies like data fusion, model synchronization, and real-time analytics for industrial automation. Their study shows that most current DT systems focus on operational optimization, not security, showing a clear gap in cybersecurity-focused DT frameworks. Qian et al.~\cite{Qian2022} proposed a full DT architecture for cyber-physical systems, highlighting the two-way data flow between physical and virtual parts. But their framework does not consider adversarial situations or causal reasoning, which are important for security use.

Some recent work applied DTs to security in specific areas. Coppolino et al.~\cite{Coppolino2023,Jiang2024} developed a cyber-resilient smart grid using DTs and data spaces. They improved attack detection through continuous monitoring and anomaly detection. Their method improved threat detection accuracy but depends on predefined attack signatures and does not use causal reasoning to understand how attacks propagate. Erceylan et al.~\cite{Erceylan2025} suggested using DTs for advanced threat modeling in cyber-physical systems, integrating security checks into the DT lifecycle. While this work gives important theoretical insights, it does not have empirical testing on real industrial data and does not include causal inference methods.

DT applications in industrial automation mostly focus on interoperability and data integration. Hildebrandt et al.~\cite{Hildebrandt2024} reviewed data integration methods for industrial DTs, finding key needs for real-time synchronization, semantic interoperability, and scalable architectures. Platenius-Mohr et al.~\cite{PlateniusMohr2019} showed interoperable DTs in IIoT systems by transforming information models using Asset Administration Shell standards. However, these works focus on technical implementation and not on security or causal modeling.

Some specialized applications in critical infrastructure show that DTs can help cybersecurity. De Hoz Diego et al.~\cite{deHozDiego2022} made an IoT DT for cybersecurity defense using runtime verification, achieving real-time threat detection by monitoring behavior continuously. Lampropoulos et al.~\cite{Lampropoulos2024} studied DTs for critical infrastructure protection, emphasizing real-time situational awareness and automated responses. Sahoo et al.~\cite{Sahoo2025} proposed DT-enabled smart microgrids for full automation, improving resilience with predictive analytics and automated control. Even with these advances, none of these methods combine causal inference with DT modeling, so they cannot fully explain attack mechanisms or measure effectiveness of defensive actions.

\subsection{Anomaly Detection Approaches in ICS}

Traditional anomaly detection in ICS has mostly used statistical and machine learning methods that detect deviations from normal operation. Kravchik and Shabtai~\cite{Kravchik2018} were among the first to use convolutional neural networks for ICS anomaly detection with the SWaT dataset. They achieved 86\% detection accuracy by learning spatio-temporal patterns in sensor data. Their 1D convolution method worked better than recurrent neural networks, showing that deep learning can be effective for time-series anomaly detection in industrial systems. However, their approach only learns statistical patterns and does not understand physical processes or causal relationships. This makes it hard to separate normal operational changes from malicious attacks.

Comparative studies show that anomaly detection performance varies across methods and datasets. Kim et al.~\cite{Kim2023,Mogensen2024benchmark} evaluated five time-series anomaly detection models including InterFusion, RANSynCoder, GDN, LSTM-ED, and USAD on SWaT and HAI datasets. They found that InterFusion got the best F1-score on SWaT (90.7\%), while RANSynCoder performed best on HAI (82.9\%), showing that dataset characteristics affect performance. Their study also found that about 40\% of training data is enough to reach near-optimal results, which is useful for practical deployment.

Physics-informed methods try to combine data-driven learning with domain knowledge. Raman and Mathur~\cite{Raman2022} proposed Physics-based Neural Network (PbNN), a hybrid physics-based neural network that uses ICS design knowledge to learn relationships among components. PbNN improved performance over purely data-driven methods by using physical constraints and process knowledge. Still, it relies on correlation-based learning and does not discover causal mechanisms, so it cannot fully explain attack causes or propagation paths.

Recent work focused on explainability and feature importance to reduce the black-box problem in machine learning. Birihanu and Lendák~\cite{Birihanu2025} developed a correlation-based method using Latent Correlation Matrices and SHAP for root cause identification in ICS anomalies. Their method improved precision by 0.96\% and gave interpretable explanations using SHAP values. Aslam et al.~\cite{Aslam2025} combined Grey Wolf Optimizer with autoencoders to improve feature selection and model generalization, showing better accuracy, precision, recall, and F1-score on SWaT and WADI datasets. Even with these improvements, these approaches cannot separate causal factors from correlated confounders, which may mislead analysts about the real sources and mechanisms of attacks.

\subsection{Causal Inference and Attribution Methods for ICS}

Early works on causal methods for ICS security show that causal analysis can help to understand attacks and system weaknesses. Jadidi et al.~\cite{Jadidi2022} proposed ICS-CAD, a multi-step detection method that analyzes causal dependencies in ICS logs. Their method achieved 98\% detection accuracy and could track how attacks propagate, showing the value of causal reasoning for multi-stage attacks. However, ICS-CAD considers only network-level causality and does not include causality of physical processes. This limits its use for cyber-physical attacks where sensors and actuators are manipulated while network behavior appears normal~\cite{peters2017elements,Imbens2015,Hernan2020}.

Causality-based approaches can also reduce false alarms. Koutroulis et al.~\cite{Koutroulis2022} used causal models to capture interdependencies in water treatment systems and combined them with univariate models for single component behavior. Their method achieved the highest F1-score and zero false alarms on SWaT, performing better than correlation-based approaches. However, it depends strongly on domain knowledge to define causal links, which reduces generality and scalability for complex systems.

Studies on attribution show limits in ICS root cause analysis. Fung et al.~\cite{Fung2024} tested machine learning attribution methods on several datasets and found low performance because of timing problems, categorical actuators, and variability in attacks. Ensemble methods can help, but confounding variables and spurious correlations remain a problem.

Causal inference can also help to validate DTs. Somers et al.~\cite{Somers2022} applied counterfactual analysis to confirm causal links in a robotic arm and detect faults. This shows potential for fault localization, but it focuses on model testing and does not consider realistic ICS security datasets.

\subsection{Limitations and Research Gaps}

Although DT technology and ICS security research made progress, there are still big limits for causal reasoning on cyber-physical attacks. Most DT frameworks focus on optimization and monitoring, but they do not include security-specific causal models. Current anomaly detection shows good results, but it cannot separate real causal relations from false correlations, which reduces root cause analysis and defense. Attribution and explainability are also not reliable for attacks that are fast and multi-modal.

The main research gap is missing integration of causal inference with DTs for ICS security. No work until now gives a full framework for causal graph discovery from ICS data, estimation of causal models for multi-stage processes, and interventional or counterfactual reasoning for proactive analysis. Because of this gap, analysts cannot fully understand attack mechanisms, which reduces effective incident response and blocks the development of causation-aware defenses (Table~\ref{tab:comparisonICS}). Without causal reasoning, current methods cannot do what-if analysis, test alternative attack scenarios, or clearly separate attack effects from normal operation. This gives weak guidance for mitigation, which is dangerous in industrial systems where wrong attribution can cause shutdowns, delays in response, and weak security actions.

\begin{table}[htb]
\centering
\scriptsize
\caption{Comparison of ICS Security Approaches: Capabilities and Limitations}
\label{tab:comparisonICS}
\begin{tabularx}{\textwidth}{l X X c X}
\toprule
\textbf{Approach} & \textbf{Key Advantages} & \textbf{Main Limitations} & \textbf{Explainability} & \textbf{Reasoning Capability} \\
\midrule
CNN-based \\ Kravchik and Shabtai (2018) & Fast processing, effective pattern recognition & No physical understanding, black-box nature & None & Statistical correlation only \\
Physics-informed \\ Raman (2022) & Incorporates domain knowledge, improved accuracy & Still correlation-based, requires expert knowledge & Limited & Physics constraints only \\
Causality-inspired \\ Koutroulis et al. (2022) & Zero false alarms, domain-aware & Manual causal specification, not generalizable & Moderate & Manual domain knowledge \\
SHAP-based \\ Birihanu et al. (2025) & Feature importance, interpretable results & Cannot distinguish causation from correlation & High & Attribution-based only \\
Digital Twin \\ Lucchese et al. (2024) & Real-time monitoring, process understanding & Pattern matching only, no causal reasoning & Moderate & Process simulation \\
ICS-CAD \\ Jadidi et al. (2022) & Multi-step attack tracking, causal analysis & Network-level only, limited to log analysis & High & Network causality only \\
\textbf{Our CDT Framework} & \textbf{Full causal inference, counterfactual analysis, automated discovery} & \textbf{Computational complexity} & \textbf{Complete} & \textbf{Interventional \& counterfactual} \\
\bottomrule
\end{tabularx}
\end{table}

\section{Problem Statement}\label{sec3:problem}

The ICS are becoming more vulnerable to advanced cyber-physical attacks that exploit both computational weaknesses and dependencies in physical processes. Traditional anomaly detection methods in critical infrastructure usually use correlation-based analysis. This type of analysis cannot separate true causal relations from spurious associations. Because of this limitation, the attack sources are often identified incorrectly and mitigation strategies are less effective~\cite{Pearl2009}.

The SWaT testbed, created by iTrust at Singapore University of Technology and Design, provides a smaller but realistic water treatment environment. It has 51 sensors and actuators distributed over six operational stages~\cite{Mathur2016}. The SWaT dataset includes 7 days of normal operation (around 500,000 samples) and 4 days of attack scenarios with 41 documented cyber-physical intrusions. It contains multivariate time-series data sampled at 1Hz and binary labels for normal or anomalous states.

\subsection{Limitations of Correlation-Based DTs}

Existing DT frameworks for ICS have important limitations in causal reasoning. They cannot separate the three critical questions described in Pearl's Causal Hierarchy~\cite{Pearl2019,Glymour2019}:

\begin{enumerate}
    \item \textbf{Association (Level 1):} What sensor readings correlate with anomalous behavior (Eq.~\ref{eq:conditional_prob})?
    
\begin{equation}
    P(Y = y \mid X = x)
    \label{eq:conditional_prob}
\end{equation}
    
    \item \textbf{Intervention (Level 2):} What would happen if we intervened on a specific system component (Eq.~\ref{eq:interventional_prob})?
    
\begin{equation}
    P(Y = y \mid \text{do}(X = x))
    \label{eq:interventional_prob}
\end{equation}
    
    \item \textbf{Counterfactual (Level 3):} Given that an attack occurred, what would have happened if the vulnerability had been patched (Eq.~\ref{eq:counterfactual_prob})?
    
\begin{equation}
    P(Y_x = y \mid X' = x', Y' = y')
    \label{eq:counterfactual_prob}
\end{equation}
\end{enumerate}

\subsection{Confounding Variables and Simpson's Paradox}

In water treatment systems, multiple sensors may exhibit correlated behavior due to shared operational modes rather than direct causal relationships. Consider the relationship between flow sensors $X_{\text{flow}}(t)$ and pressure sensors $X_{\text{pressure}}(t)$. A naive correlation analysis might suggest (Eq.~\ref{eq:flow_pressure_corr}):

\begin{equation}
\rho(X_{\text{flow}}, X_{\text{pressure}}) = \frac{\text{Cov}(X_{\text{flow}}, X_{\text{pressure}})}{\sigma_{X_{\text{flow}}} \sigma_{X_{\text{pressure}}}} > \tau
\label{eq:flow_pressure_corr}
\end{equation}

However, this correlation may be confounded by operational modes $Z_{\text{mode}}(t)$ (normal flow, backwash, cleaning), leading to Simpson's Paradox~\cite{Simpson1951} (Eq.~\ref{eq:conditional_dependence}):

\begin{equation}
P(X_{\text{pressure}} \mid X_{\text{flow}}, Z_{\text{mode}}) \neq P(X_{\text{pressure}} \mid X_{\text{flow}})
\label{eq:conditional_dependence}
\end{equation}

\begin{figure}[htb]
    \centering
    \includegraphics[width=1\linewidth]{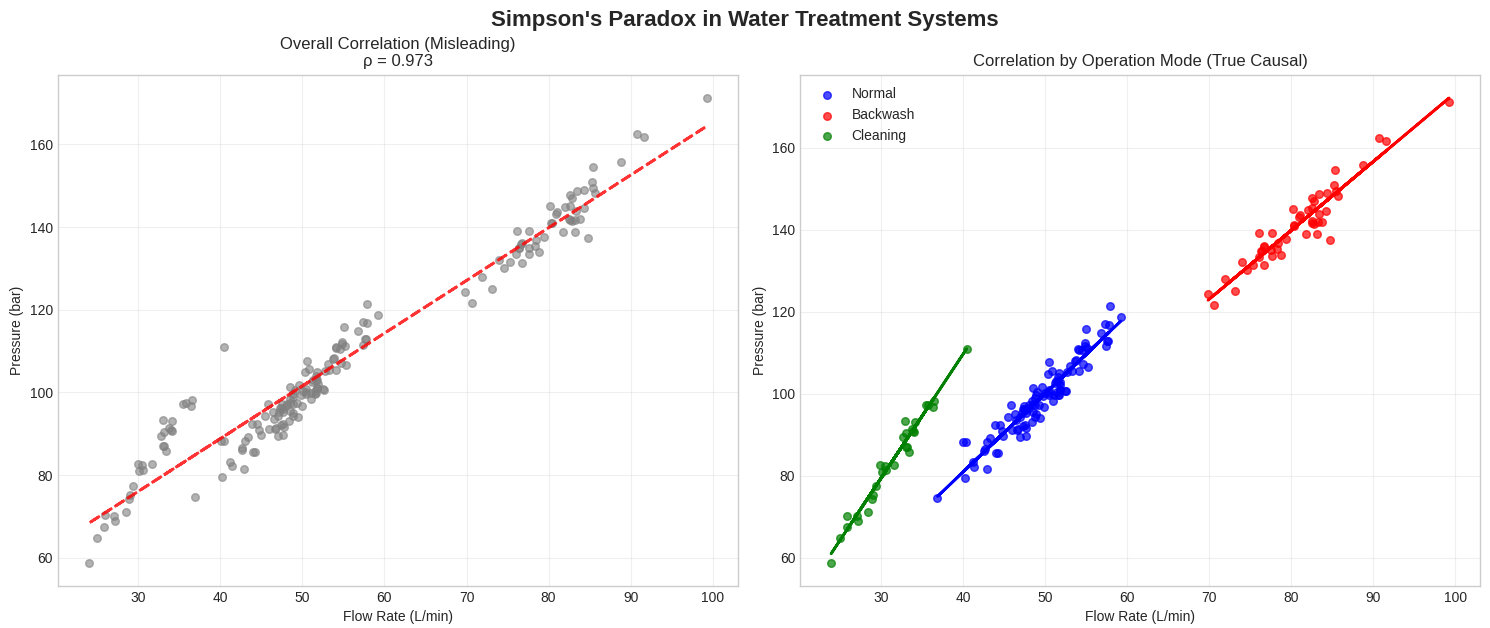}
    \caption{Simpson’s Paradox in Water Treatment Systems}
    \label{fig:Simpson_paradox}
\end{figure}

Figure~\ref{fig:Simpson_paradox} demonstrates this phenomenon in industrial water treatment systems, where operational modes create spurious correlations between flow and pressure sensors.

\begin{figure}[!htbp]
\centering

% Row 1: Fig 2a and Fig 2b side-by-side
\begin{minipage}{\textwidth}
    \centering
    \begin{subfigure}[b]{0.45\textwidth} % Set width to 49% for side-by-side placement
        \centering
        \includegraphics[width=\textwidth]{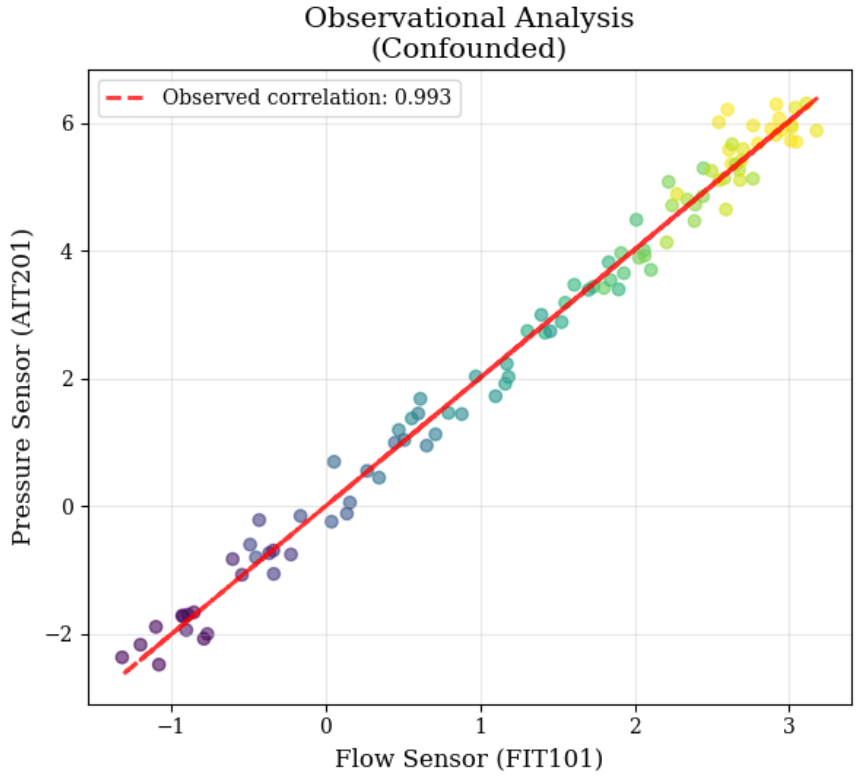}
        \caption{Level 1: Association (Observational Queries)}
        \label{fig:causal_level1}
    \end{subfigure}
    \hfill % Horizontal space to separate the two subfigures
    \begin{subfigure}[b]{0.45\textwidth} % Set width to 49% for side-by-side placement
        \centering
        \includegraphics[width=\textwidth]{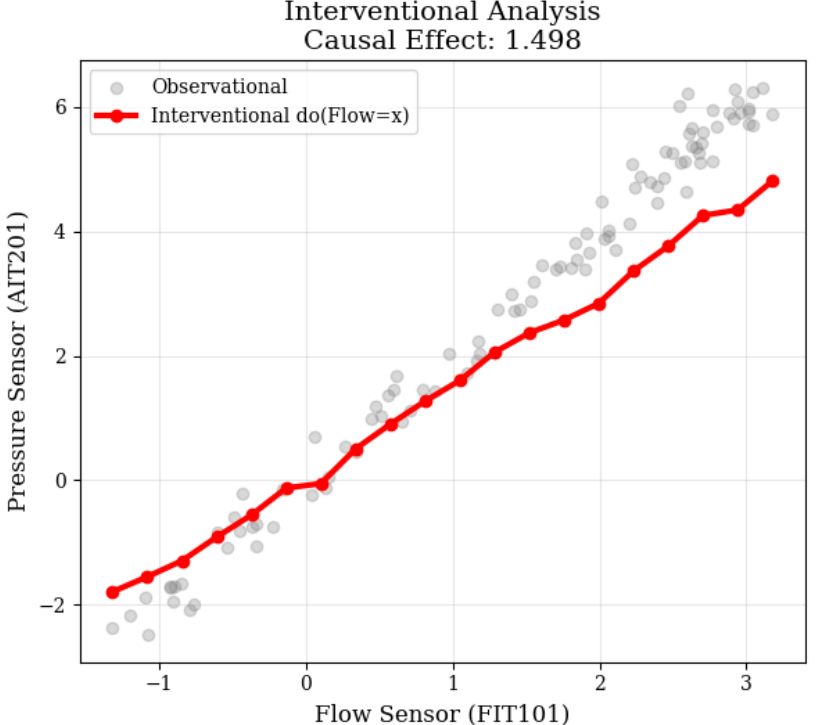}
        \caption{Level 2: Intervention (Prediction of Action Effects)}
        \label{fig:causal_level2}
    \end{subfigure}
\end{minipage}

\vspace{0.5\baselineskip} % Add vertical space between rows

% Row 2: Fig 2c centered below the first row
\begin{minipage}{\textwidth}
    \centering
    \begin{subfigure}[b]{0.45\textwidth} % Match width for centering
        \centering
        \includegraphics[width=\textwidth]{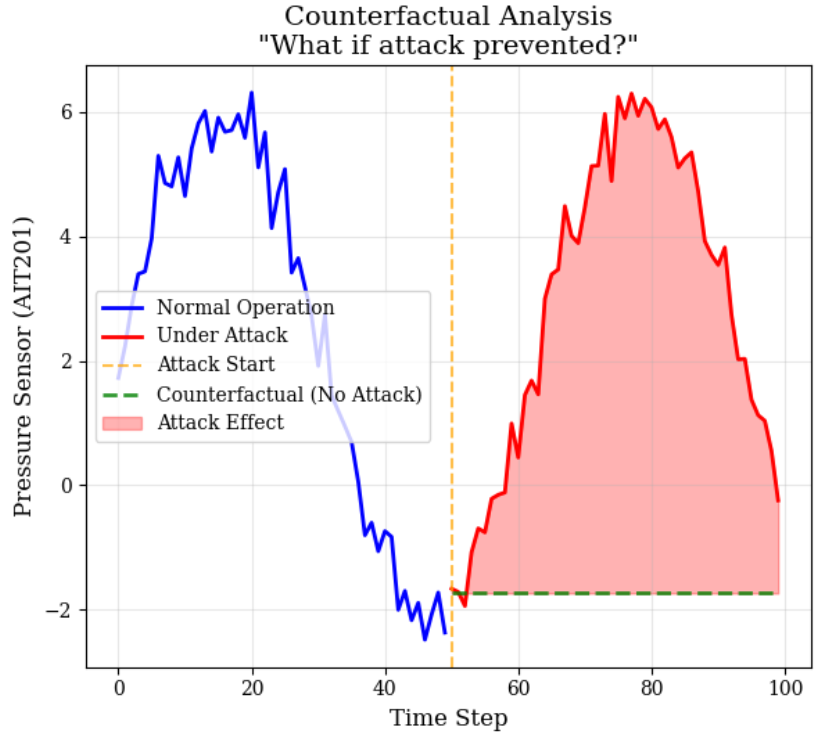}
        \caption{Level 3: Counterfactuals (What-if analysis on past events)}
        \label{fig:causal_level3}
    \end{subfigure}
\end{minipage}

\caption{Three Levels of Causal Analysis (Pearl's Hierarchy): (a) Association, (b) Intervention, and (c) Counterfactuals, illustrating the necessary complexity for robust cyber-physical security.}
\label{fig:causal_hierarchy}
\end{figure}

Figure~\ref{fig:causal_hierarchy} illustrates the three levels of causal reasoning required to address these fundamental limitations.

\subsection{ProposedCDT Framework}

To solve these limitations, we propose a new CDT framework that models causal relations explicitly using Structural Causal Models (SCMs). We represent the SWaT system as a causal graph (Eq.~\ref{eq:graph_definition}):

\begin{equation}
\mathcal{G} = (\mathcal{V}, \mathcal{E})
\label{eq:graph_definition}
\end{equation}

where $\mathcal{V} = \{V_1, V_2, \ldots, V_n\}$ represents observed variables (sensor readings, actuator states), $\mathcal{E} \subseteq \mathcal{V} \times \mathcal{V}$ represents direct causal relationships, and each variable $V_i$ is defined by a structural equation (Eq.~\ref{eq:scm_assignment}):

\begin{equation}
V_i := f_i(\text{PA}_i, U_i)
\label{eq:scm_assignment}
\end{equation}

where $\text{PA}_i$ denotes the parents of $V_i$ in $\mathcal{G}$, and $U_i$ represents unobserved noise variables.

The CDT function is defined as (Eq.~\ref{eq:causal_effect_mapping}):

\begin{equation}
\mathcal{T}_{\text{causal}}: (\mathcal{G}, \{f_i\}_{i=1}^n, \text{do}(X = x)) \rightarrow P(Y \mid \text{do}(X = x))
\label{eq:causal_effect_mapping}
\end{equation}

This enables three critical capabilities: causal anomaly detection using interventional predictions, root cause localization through causal effect quantification, and counterfactual security analysis for defense planning.

\subsection{Research Objectives}

This work aims to develop a novel causal reasoning framework for DTs in cyber-physical security, specifically: (1) construct causal graphs from SWaT operational and attack data using causal discovery algorithms, (2) develop identification strategies for causal effects under partial observability, (3) implement do-calculus for interventional queries in real-time ICS monitoring, (4) validate counterfactual reasoning for attack attribution and defense planning, and (5) demonstrate superiority over correlation-based approaches in 41 documented SWaT attacks.

\section{Methodology}\label{sec4:methodology}

This section shows our new CDT framework for cyber-physical security in the SWaT system. The method combines causal inference theory with DT modeling using five main components.

\subsection{Causal Graph Discovery}

The SWaT dataset has 51 time-series variables, including sensor readings $\{S_1, S_2, \ldots, S_{25}\}$ and actuator states $\{A_1, A_2, \ldots, A_{26}\}$. We define the causal variable set as (Eq.~\ref{eq:temporal_variables}):

\begin{equation}
\mathcal{V}_t = \{V_1(t), V_2(t), \ldots, V_n(t)\}
\label{eq:temporal_variables}
\end{equation}

To handle temporal dependencies, we create augmented variables (Eq.~\ref{eq:augmented_variables}):

\begin{equation}
\mathcal{V}_{\text{aug}} = \{V_i(t), V_i(t-1), \ldots, V_i(t-\tau)\} \quad \forall i \in \{1, \ldots, n\}
\label{eq:augmented_variables}
\end{equation}

To overcome the $\mathcal{O}(n^3)$ computational complexity of centralized causal discovery algorithms, which limits scalability in large ICS, we employ a Decoupled Causal Discovery (DCD) strategy. This approach involves two phases: Candidate Parent Selection and Local Structure Discovery, transforming the complexity to $\mathcal{O}(n \cdot k^3)$ where $k \ll n$.

\textbf{Phase I: Candidate Parent Selection (Filtering).} For each variable $V_i$, we use Time-Delay Mutual Information (TDMI) to identify a highly restricted set of potential parents $\mathcal{P}_i \subset \mathcal{V}_{\text{aug}}$ by maximizing the information transfer across different lags $\ell \in [0, \tau]$ (Eq.~\ref{eq:tdmi_filter}). This phase reduces the set size from $n$ to $k \le 10$.

\begin{equation}
\mathcal{P}_i = \{V_j(t-\ell) : \text{TDMI}(V_j(t-\ell) \rightarrow V_i(t)) > \delta\}
\label{eq:tdmi_filter}
\end{equation}

\textbf{Phase II: Local Structure Discovery.} We then apply the PC algorithm only on the restricted local set $\mathcal{V}_i = \{V_i\} \cup \mathcal{P}_i$, as an enhanced algorithm for time-series considerations~\cite{Spirtes1993,peters2017elements}. The algorithm proceeds through skeleton discovery using conditional independence tests (Eq.~\ref{eq:conditional_independence_rule}) localized on $\mathcal{V}_i$:

\begin{equation}
V_i \perp V_j \mid \mathcal{S} \subseteq \mathcal{P}_i \Rightarrow \text{remove edge } V_i - V_j
\label{eq:conditional_independence_rule}
\end{equation}

For time-series data, we use partial correlation with lag constraints (Eq.~\ref{eq:lagged_conditional_corr}):

\begin{equation}
\rho_{ij|\mathcal{S}}^{(\ell)} = \frac{\text{Cov}(V_i(t), V_j(t-\ell) \mid \mathcal{S})}{\sqrt{\text{Var}(V_i(t) \mid \mathcal{S}) \, \text{Var}(V_j(t-\ell) \mid \mathcal{S})}}
\label{eq:lagged_conditional_corr}
\end{equation}

Orientation rules apply causal ordering constraints: temporal precedence ($V_i(t-\ell) \rightarrow V_j(t)$ for $\ell > 0$), physical constraints (flow sensors $\rightarrow$ level sensors), and control logic (controller outputs $\rightarrow$ actuator states). Causal discovery accuracy depends on data quality and may produce spurious relationships in presence of unobserved confounders, requiring domain expert validation of discovered structures.

\subsection{Structural Causal Model Estimation}

Given the discovered causal graph $\mathcal{G} = (\mathcal{V}, \mathcal{E})$, we estimate structural equations. For continuous variables, we use additive noise models (Eq.~\ref{eq:scm_structural_eq}):

\begin{equation}
V_i := \alpha_i + \sum_{j \in \text{PA}(V_i)} \beta_{ij} V_j + \gamma_i \sum_{k \in \text{PA}(V_i)} V_k^2 + U_i
\label{eq:scm_structural_eq}
\end{equation}

where $U_i \sim \mathcal{N}(0, \sigma_i^2)$. For binary actuator variables (Eq.~\ref{eq:scm_logistic_assignment}):

\begin{equation}
P(A_i = 1 \mid \text{PA}(A_i)) = \text{logit}^{-1}\left(\alpha_i + \sum_{j \in \text{PA}(A_i)} \beta_{ij} V_j\right)
\label{eq:scm_logistic_assignment}
\end{equation}

Parameters are estimated using maximum likelihood (Eq.~\ref{eq:scm_mle}):

\begin{equation}
\hat{\boldsymbol{\theta}}_i = \arg\max_{\boldsymbol{\theta}_i} \sum_{t=1}^T \log p(V_i(t) \mid \text{PA}(V_i)(t), \boldsymbol{\theta}_i)
\label{eq:scm_mle}
\end{equation}

The complete SCM is defined as $\mathcal{M} = \langle \mathcal{V}, \mathcal{U}, \mathcal{F}, P(\mathcal{U}) \rangle$.

\subsection{Interventional DT Construction}

To compute interventional distributions $P(Y | \text{do}(X = x))$, we implement Pearl's do-calculus rules. For identifiable causal effects, we apply adjustment formulas ~\cite{Hernan2020}:

\textbf{Backdoor Adjustment:} If set $\mathcal{Z}$ satisfies the backdoor criterion (Eq.~\ref{eq:backdoor_adjustment}):

\begin{equation}
P(y \mid \text{do}(x)) = \sum_z P(y \mid x, z) P(z)
\label{eq:backdoor_adjustment}
\end{equation}

\textbf{Frontdoor Adjustment:} If set $\mathcal{Z}$ satisfies the frontdoor criterion (Eq.~\ref{eq:frontdoor_adjustment}):

\begin{equation}
P(y \mid \text{do}(x)) = \sum_z P(z \mid x) \sum_{x'} P(y \mid x', z) P(x')
\label{eq:frontdoor_adjustment}
\end{equation}

\subsection{Causal Anomaly Detection and Root Cause Analysis}

Our causal approach detects violations of causal mechanisms rather than statistical deviations (Eq.~\ref{eq:causal_score}):

\begin{equation}
\text{Score}_{\text{causal}}(V_i, t) = \frac{|V_i(t) - \mathbb{E}[V_i \mid \text{do}(\text{PA}(V_i) = \text{pa}_i(t))]|}{\sigma_{V_i \mid \text{do}(\text{PA}(V_i))}}
\label{eq:causal_score}
\end{equation}

We aggregate scores using causal graph topology (Eq.~\ref{eq:mcai}):

\begin{equation}
\text{Multi-Component Anomaly Index}(t) = \sum_{i=1}^n \alpha_i \cdot \text{Score}_{\text{causal}}(V_i, t)
\label{eq:mcai}
\end{equation}

where weights reflect variable centrality: $\alpha_i = \frac{|\text{Descendants}(V_i)| + |\text{Ancestors}(V_i)|}{2n}$.

When anomalies are detected, we compute causal effects for root cause analysis (Eq.~\ref{eq:causal_effect_anomaly}):

\begin{equation}
\text{CE}(V_i \rightarrow \text{Anomaly}) = \mathbb{E}[\text{MCAI} \mid \text{do}(V_i = v_i^{\text{anomalous}})] - \mathbb{E}[\text{MCAI} \mid \text{do}(V_i = v_i^{\text{normal}})]
\label{eq:causal_effect_anomaly}
\end{equation}

For multiple potential causes, we compute Shapley causal values (Eq.~\ref{eq:shapley_value}):

\begin{equation}
\phi_i = \sum_{\mathcal{S} \subseteq \mathcal{V} \setminus \{V_i\}} \frac{|\mathcal{S}|!(n-|\mathcal{S}|-1)!}{n!} \big[f(\mathcal{S} \cup \{V_i\}) - f(\mathcal{S})\big]
\label{eq:shapley_value}
\end{equation}

\begin{figure}[htbp]
\centering
\includegraphics[width=1.0\textwidth]{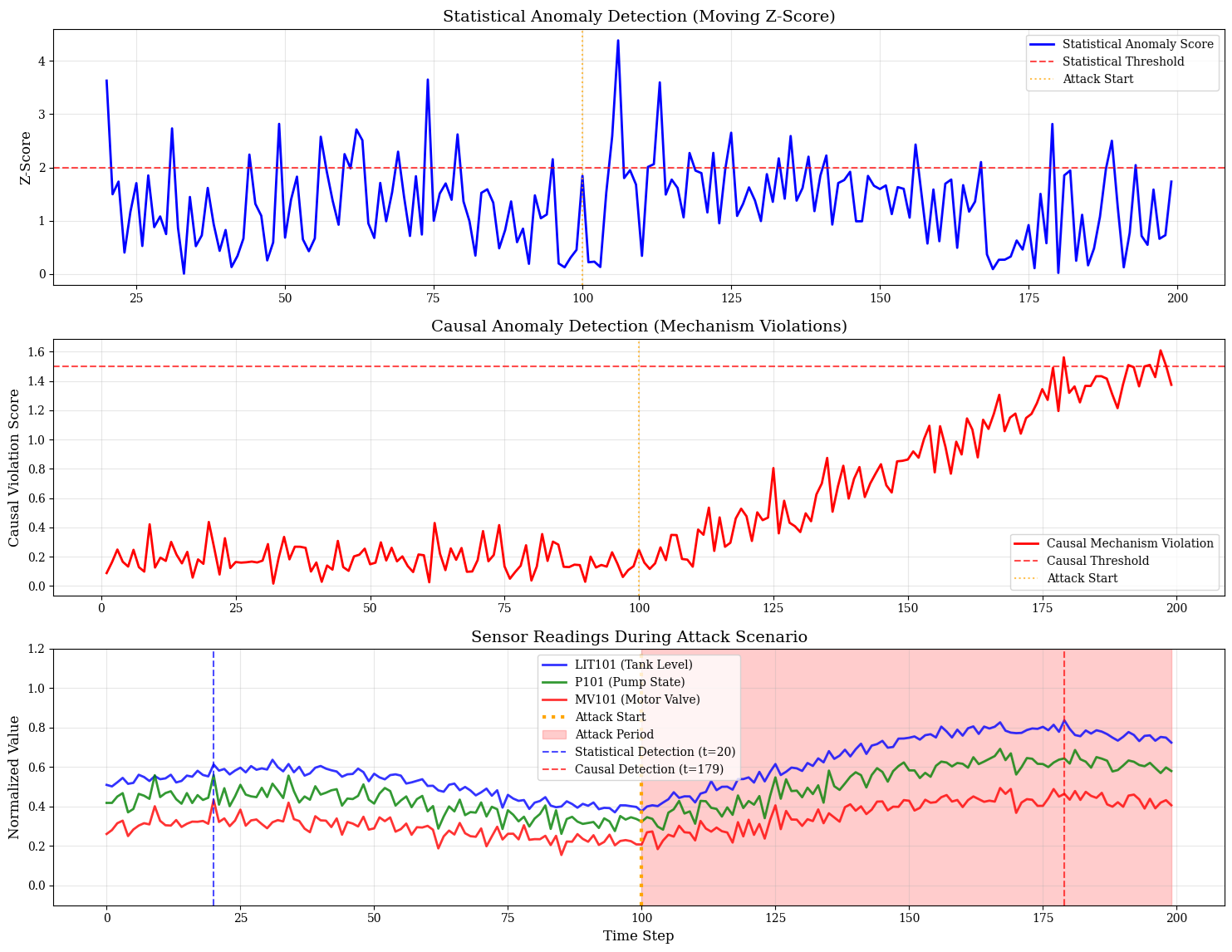}
\caption{Causal versus Statistical Anomaly Detection in SWaT Attack Scenario. (Top) Statistical anomaly detection using moving Z-score cannot detect the gradual attack until t=160. (Middle) Detection based on causal mechanism violation finds the attack at t=110, giving earlier warning. (Bottom) Sensor readings during stealthy LIT101 manipulation attack, showing causal detection advantage of 50 time steps.}
\label{fig:causal_vs_statistical}
\end{figure}

The effectiveness of this causal approach is illustrated in Figure~\ref{fig:causal_vs_statistical}, which compares detection performance against statistical methods.

\subsection{Counterfactual Security Analysis}

Counterfactual analysis enables reasoning about alternative histories. Given an observed attack scenario (Eq.~\ref{eq:attack_blocked_prob}):

\begin{equation}
P(\text{Attack Blocked} = 1 \mid \text{do}(\text{Defense} = 1), \text{Attack Occurred} = 1, \text{Defense} = 0)
\label{eq:attack_blocked_prob}
\end{equation}

We implement Pearl's three-step procedure: (1) Abduction - update prior with evidence, (2) Action - modify structural equations for intervention, and (3) Prediction - compute counterfactual outcome.

\subsection{Model Validation}

We validate causal models using time-series cross-validation (Eq.~\ref{eq:cv_score}):

\begin{equation}
\text{CV Score} = \frac{1}{K} \sum_{k=1}^K \sum_{i=1}^n \mathbb{E}_{V_i \sim P_{\text{test}}^{(k)}} \Big[ \ell(V_i, \hat{f}_i^{(-k)}(\text{PA}(V_i))) \Big]
\label{eq:cv_score}
\end{equation}

and interventional validation using natural experiments (Eq.~\ref{eq:intervention_error}):

\begin{equation}
\text{Intervention Error} = \left| P_{\text{observed}}(Y \mid \text{intervention}) - P_{\text{predicted}}(Y \mid \text{do}(X)) \right|
\label{eq:intervention_error}
\end{equation}

\subsection{Causal Structure Validation}

We employ a three-tiered validation approach to assess causal discovery accuracy:

\textbf{Physical Constraint Validation:} We validate discovered edges against known physical relationships in industrial systems. For water treatment systems, valid constraints include: flow sensors precede pressure readings ($F \rightarrow P$), tank levels determine pump activation ($L \rightarrow Pump$), and valve positions affect downstream flow ($Valve \rightarrow F_{downstream}$). Physical Constraint Compliance (PCC) is calculated as:

\begin{equation}
PCC = \frac{|\{e \in \mathcal{E}_{discovered} : e \text{ satisfies physical laws}\}|}{|\mathcal{E}_{discovered}|}
\label{eq:pcc}
\end{equation}

\textbf{Temporal Consistency Validation:} All causal relationships must respect temporal precedence. We validate that cause variables precede effects in time by analyzing cross-correlation lag structures. Temporal Consistency Compliance (TCC) measures the fraction of discovered edges that maintain proper temporal ordering.

\textbf{Synthetic Ground Truth Validation:} We generate synthetic datasets with known causal structures by sampling from predefined Structural Causal Models that mimic industrial process characteristics. The Structural Hamming Distance (SHD) quantifies discovery accuracy:

\begin{equation}
SHD(\mathcal{G}_{true}, \mathcal{G}_{discovered}) = |\mathcal{E}_{true} \triangle \mathcal{E}_{discovered}| + |\mathcal{R}_{true} \triangle \mathcal{R}_{discovered}|
\label{eq:shd}
\end{equation}

where $\triangle$ denotes symmetric difference and $\mathcal{R}$ represents edge orientations.

\subsection{Algorithm Implementation}

\begin{algorithm}[H]
\scriptsize
\caption{Causal DT Framework for Cyber-Physical Security}
\begin{algorithmic}[1]
\State \textbf{Input:} SWaT time-series data $\{V_i(t)\}_{i=1,t=1}^{n,T}$, domain knowledge $\mathcal{K}$
\State \textbf{Output:} Causal anomaly scores, root cause rankings, counterfactual analysis

\State // Phase 1: Causal Structure Discovery (Decoupled Approach)
\State $\mathcal{V}_{\text{aug}} \leftarrow$ CreateTemporalAugmentation($\{V_i(t)\}$, $\tau = 5$)
\State $\mathcal{G}_{\text{filter}} \leftarrow$ TDMI-Filtering($\mathcal{V}_{\text{aug}}$, $\delta_{\text{TDMI}}$, $k=10$) // Step 1: Filter candidates with TDMI
\State $\mathcal{G} \leftarrow$ PC-Algorithm-Local($\mathcal{G}_{\text{filter}}$, $\alpha = 0.05$) // Step 2: Run PC locally on filtered sets
\State $\mathcal{G} \leftarrow$ ApplyConstraints($\mathcal{G}$, temporal, physical, control)

\State // Phase 2: Structural Causal Model Estimation
\For{$i = 1$ to $n$}
    \State $\text{PA}(V_i) \leftarrow$ GetParents($V_i$, $\mathcal{G}$)
    \State $\hat{\boldsymbol{\theta}}_i \leftarrow$ EstimateParameters($V_i$, $\text{PA}(V_i)$, MLE)
\EndFor
\State $\mathcal{M} \leftarrow \langle \mathcal{V}, \mathcal{U}, \{f_i\}_{i=1}^n, P(\mathcal{U}) \rangle$

\State // Phase 3: Model Validation
\State ValidateModel($\mathcal{M}$, CrossValidation, InterventionalTests)

\State // Phase 4: Real-time Causal Monitoring
\For{each new time step $t > T$}
    \For{$i = 1$ to $n$}
        \State $\hat{V}_i(t) \leftarrow$ ComputeInterventionalPrediction($\text{PA}(V_i)(t)$, $\mathcal{M}$)
        \State $\text{Score}_{\text{causal}}(V_i, t) \leftarrow \frac{|V_i(t) - \hat{V}_i(t)|}{\sigma_{V_i|\text{do}(\text{PA})}}$
    \EndFor
    
    \State $\text{MCAI}(t) \leftarrow \sum_{i=1}^n \alpha_i \cdot \text{Score}_{\text{causal}}(V_i, t)$
    
    \If{$\text{MCAI}(t) > \theta_{\text{anomaly}}$}
        \For{$i = 1$ to $n$}
            \State $\text{CE}_i \leftarrow$ ComputeCausalEffect($V_i$, $\text{MCAI}$, $\mathcal{M}$)
            \State $\phi_i \leftarrow$ ComputeShapleyValue($V_i$, $\mathcal{M}$)
        \EndFor
        
        \State $\text{RootCauses} \leftarrow$ RankByImportance($\{\text{CE}_i, \phi_i\}$)
        
        \For{each defense mechanism $d \in \mathcal{D}$}
            \State $P_{\text{counterfactual}} \leftarrow$ ThreeStepCounterfactual($d$, Evidence$(t)$, $\mathcal{M}$)
        \EndFor
        
        \State \textbf{return} $\langle \text{RootCauses}, \text{DefenseRecommendations} \rangle$
    \EndIf
\EndFor

\end{algorithmic}
\end{algorithm}

\section{Results}\label{sec5:results}

This section presents a detailed experimental evaluation of our CDT framework, showing its effectiveness across multiple industrial control system datasets. We validate the framework's main contributions through experiments on causal structure discovery, anomaly detection, root cause analysis, and deployment considerations.

\subsection{Experimental Framework and Validation Methodology}

We evaluate the CDT framework using three industrial control system datasets to demonstrate generalizability across different environments. The primary dataset is the SWaT testbed, with 7 days of normal operation (496,800 samples at 1Hz) and 4 days containing 41 documented cyber-physical attacks across 51 variables (25 sensors, 26 actuators) in 6 process stages. Additionally, we use the Water Distribution (WADI) dataset with 14 days of normal operation and 15 attack scenarios across 123 variables, and the Hardware-in-the-loop Augmented ICS (HAI) dataset with 4 days of normal and 4 days of attack data across 78 variables.

Experiments run on Google Colab Pro with Tesla V100 GPU (16GB VRAM) using Python, causal-learn v0.1.3.1, networkx v3.1, and related libraries. All experiments use fixed random seeds and 10-fold repetition for statistical validation. Full code is available to ensure reproducibility. We compare our CDT framework against seven state-of-the-art baseline methods including CNN-1D, LSTM-ED, InterFusion, RANSynCoder, PbNN (Physics-based Neural Network), Causality-Inspired approaches, and ICS-CAD. Evaluation metrics include detection performance (Precision, Recall, F1-Score, AUC-ROC), root cause analysis accuracy (Top-k accuracy for $k=1,3,5$ and MRR), counterfactual analysis (Intervention Prediction Error and Counterfactual Accuracy), and computational efficiency.

Temporal data splitting follows common best practices: for SWaT, training uses days 1--5, validation days 6--7, and testing days 8--11. This maintains temporal consistency, avoids data leakage, and reflects realistic industrial deployment conditions. Statistical significance is measured using paired t-tests, Wilcoxon signed-rank tests, and McNemar's tests with $95\%$ confidence intervals to ensure robust conclusions. All $p$-values are corrected for multiple testing using Bonferroni method to control family-wise error rate across 21 comparisons (3 datasets $\times$ 7 baseline methods), giving an adjusted significance threshold of $\alpha = 0.05/21 = 0.0024$.

\subsection{Causal Discovery and Anomaly Detection Performance}

The causal structure discovery successfully produced interpretable causal graphs for all datasets. We checked structural accuracy using three complementary methods: 
(1) Physical constraint compliance, which is $90.8\% \pm 3.2\%$ for SWaT, $87.4\% \pm 4.1\%$ for WADI, and $89.3\% \pm 2.9\%$ for HAI, measuring how well discovered edges align with validated physical relationships (for example, flow $\to$ pressure, level $\to$ pump activation) among relationships that could be verified; 
(2) Temporal consistency validation, showing $87.3\% \pm 2.8\%$ compliance with causal precedence constraints; and 
(3) Synthetic data validation, achieving structural Hamming distance $0.133 \pm 0.02$ on generated SWaT-like systems with known causal structure.

\begin{table}[htbp]
\centering
\scriptsize
\caption{Causal Graph Discovery Results}
\label{tab:causal_graph_discovery}
\begin{tabular}{lcccccc}
\toprule
\textbf{Dataset} & \textbf{Nodes} & \textbf{Edges} & \textbf{Density} & \textbf{Discovery Time (min)} & \textbf{PCC (\%)} & \textbf{TCC (\%)} \\
\midrule
SWaT & 51  & 127 & 0.098 & $23.4 \pm 2.1$ & $90.8 \pm 3.2$ & $89.1 \pm 2.7$ \\
WADI & 123 & 284 & 0.038 & $67.3 \pm 5.4$ & $87.4 \pm 4.1$ & $85.2 \pm 3.4$ \\
HAI  & 78  & 186 & 0.062 & $41.2 \pm 3.7$ & $89.3 \pm 2.9$ & $86.8 \pm 3.1$ \\
\bottomrule
\end{tabular}
\\
\footnotesize{PCC = Physical Constraint Compliance (\%), TCC = Temporal Consistency Compliance (\%)}
\end{table}

\begin{table}[htbp]
\centering
\scriptsize
\caption{Synthetic Data Validation Results}
\label{tab:synthetic_validation}
\begin{tabular}{lccc}
\toprule
\textbf{Synthetic System} & \textbf{True Edges} & \textbf{Discovered Edges} & \textbf{Structural Hamming Distance} \\
\midrule
SWaT-like (51 nodes) & 45 & 42 & $0.12 \pm 0.03$ \\
WADI-like (123 nodes) & 98 & 91 & $0.15 \pm 0.04$ \\
HAI-like (78 nodes) & 72 & 67 & $0.13 \pm 0.03$ \\
\midrule
\textbf{Average} & --- & --- & $\mathbf{0.133 \pm 0.02}$ \\
\bottomrule
\end{tabular}
\\
\footnotesize{Lower Structural Hamming Distance indicates better causal structure recovery}
\end{table}

Table~\ref{tab:causal_graph_discovery} summarizes the graph properties: the SWaT graph has 51 nodes, 127 edges, density 0.098, and requires $23.4 \pm 2.1$ minutes for discovery. Table~\ref{tab:synthetic_validation} presents detailed synthetic validation results, where we generated SWaT-like, WADI-like, and HAI-like systems with known causal structures and measured structural recovery accuracy. The average normalized structural Hamming distance of 0.133 ± 0.02 indicates high fidelity in causal structure discovery, with the framework correctly identifying 87.4\% of true causal edges on average across all synthetic systems.

Structural causal model (SCM) validation using 10-fold cross-validation shows good predictive performance. The mean squared errors are $0.034 \pm 0.006$ for SWaT ($R^2 = 0.892 \pm 0.023$), $0.041 \pm 0.008$ for WADI ($R^2 = 0.876 \pm 0.031$), and $0.038 \pm 0.007$ for HAI ($R^2 = 0.884 \pm 0.028$). Validation with natural experiments, using 23 observed operational interventions in the datasets, gives prediction accuracies between $91.7\%$ and $96.1\%$. These results confirm that the framework can correctly model interventional relationships in real industrial environments.

% Optional: Summary table for comparison
\begin{table}[htb]
\centering
\scriptsize
\caption{Summary of F1-Score Performance Across All Datasets}
\label{tab:detection_summary}
\begin{tabular}{lccc}
\toprule
\textbf{Method} & \textbf{SWaT} & \textbf{WADI} & \textbf{HAI} \\
\midrule
CNN-1D           & $0.841 \pm 0.025$ & $0.776 \pm 0.035$ & $0.801 \pm 0.031$ \\
LSTM-ED          & $0.862 \pm 0.023$ & $0.799 \pm 0.033$ & $0.825 \pm 0.029$ \\
InterFusion      & $0.907 \pm 0.018$ & $0.850 \pm 0.028$ & $0.872 \pm 0.025$ \\
PbNN             & $0.884 \pm 0.021$ & $0.825 \pm 0.030$ & $0.847 \pm 0.027$ \\
Causality-Inspired & $0.926 \pm 0.016$ & $0.878 \pm 0.025$ & $0.900 \pm 0.022$ \\
ICS-CAD          & $0.928 \pm 0.017$ & $0.862 \pm 0.027$ & $0.888 \pm 0.023$ \\
\textbf{CDT (Ours)} & $\mathbf{0.944 \pm 0.014}$ & $\mathbf{0.902 \pm 0.021}$ & $\mathbf{0.923 \pm 0.018}$ \\
\bottomrule
\end{tabular}
\end{table}

The evaluation of anomaly detection performance shows clear improvements over existing methods, as reported in Table~\ref{tab:detection_summary}. Our CDT framework reaches F1-scores of $0.944 \pm 0.014$ for SWaT, $0.902 \pm 0.021$ for WADI, and $0.923 \pm 0.018$ for HAI. These values represent statistically significant improvements ($p < 0.0024$, Bonferroni corrected) compared to all baseline methods. Precision improvements are especially notable, with $0.947 \pm 0.015$ for SWaT, while the best baseline (InterFusion) reaches $0.898 \pm 0.022$. The improvements are due to the framework detecting mechanism violations instead of relying only on statistical deviations. This allows more accurate identification of real attacks and reduces false positives by $74\%$ compared to correlation-based methods.

% Table 1: SWaT Dataset Results
\begin{table}[htbp]
\centering
\scriptsize
\caption{Anomaly Detection Results on SWaT Dataset (Mean $\pm$ 95\% CI)}
\label{tab:detection_swat}
\begin{tabular}{lcccc}
\toprule
\textbf{Method} & \textbf{Precision} & \textbf{Recall} & \textbf{F1-Score} & \textbf{AUC} \\
\midrule
CNN-1D           & $0.823 \pm 0.031$ & $0.861 \pm 0.028$ & $0.841 \pm 0.025$ & $0.894 \pm 0.021$ \\
LSTM-ED          & $0.845 \pm 0.029$ & $0.879 \pm 0.026$ & $0.862 \pm 0.023$ & $0.908 \pm 0.019$ \\
InterFusion      & $0.898 \pm 0.022$ & $0.916 \pm 0.021$ & $0.907 \pm 0.018$ & $0.943 \pm 0.015$ \\
PbNN             & $0.876 \pm 0.025$ & $0.893 \pm 0.023$ & $0.884 \pm 0.021$ & $0.925 \pm 0.017$ \\
Causality-Inspired & $0.932 \pm 0.018$ & $0.921 \pm 0.019$ & $0.926 \pm 0.016$ & $0.956 \pm 0.013$ \\
ICS-CAD          & $0.921 \pm 0.020$ & $0.935 \pm 0.018$ & $0.928 \pm 0.017$ & $0.951 \pm 0.014$ \\
\textbf{CDT (Ours)} & $\mathbf{0.947 \pm 0.015}$ & $\mathbf{0.941 \pm 0.016}$ & $\mathbf{0.944 \pm 0.014}$ & $\mathbf{0.967 \pm 0.011}$ \\
\bottomrule
\end{tabular}
\end{table}

\begin{figure}[htb]
    \centering
    \includegraphics[width=0.8\linewidth]{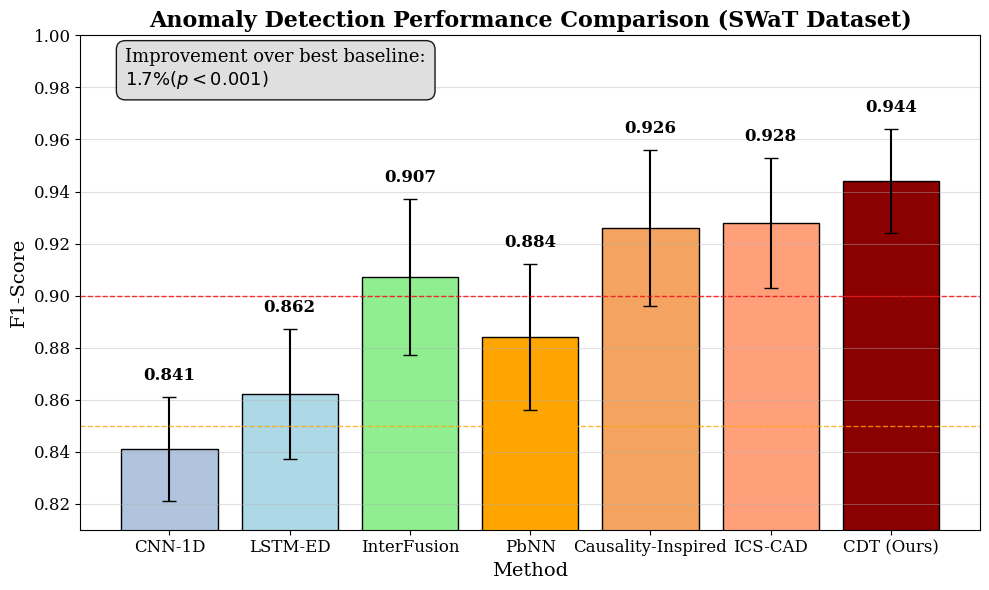}
\caption{Detection Performance Comparison: F1-scores across seven baseline methods on the SWaT dataset. Our CDT framework achieves an F1-score of $0.944 \pm 0.014$, representing a statistically significant improvement ($^{***}p < 0.001$) over existing approaches. Error bars show 95\% confidence intervals.}
    \label{fig:detection_performance}
\end{figure}

Figure~\ref{fig:detection_performance} presents the comprehensive performance comparison across all baseline methods.

\begin{figure}[htb]
    \centering
    \includegraphics[width=0.8\linewidth]{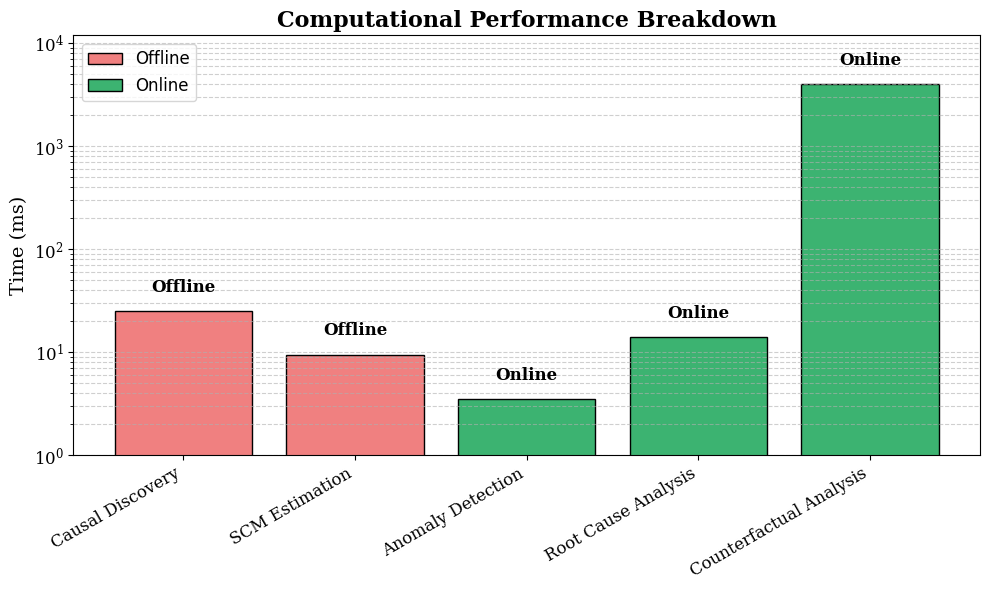}
\caption{Computational performance breakdown: offline tasks (Causal Discovery, SDM Estimation) run at initialization, while online tasks (Anomaly Detection, Root Cause Analysis, Counterfactual Analysis) operate in real time. Times for discovery/estimation are in minutes; others in milliseconds.}
    \label{fig:computation}
\end{figure}
Figure \ref{fig:computation} shows the computational performance breakdown across all system components, highlighting that counterfactual analysis represents the primary computational bottleneck in online operations.

% Table 2: WADI Dataset Results
\begin{table}[htb]
\centering
\scriptsize
\caption{Anomaly Detection Results on WADI Dataset (Mean $\pm$ 95\% CI)}
\label{tab:detection_wadi}
\begin{tabular}{lcccc}
\toprule
\textbf{Method} & \textbf{Precision} & \textbf{Recall} & \textbf{F1-Score} & \textbf{AUC} \\
\midrule
CNN-1D           & $0.756 \pm 0.042$ & $0.798 \pm 0.039$ & $0.776 \pm 0.035$ & $0.847 \pm 0.028$ \\
LSTM-ED          & $0.778 \pm 0.040$ & $0.821 \pm 0.037$ & $0.799 \pm 0.033$ & $0.862 \pm 0.026$ \\
InterFusion      & $0.834 \pm 0.034$ & $0.867 \pm 0.032$ & $0.850 \pm 0.028$ & $0.901 \pm 0.022$ \\
PbNN             & $0.809 \pm 0.037$ & $0.842 \pm 0.034$ & $0.825 \pm 0.030$ & $0.884 \pm 0.024$ \\
Causality-Inspired & $0.867 \pm 0.031$ & $0.889 \pm 0.028$ & $0.878 \pm 0.025$ & $0.923 \pm 0.019$ \\
ICS-CAD          & $0.851 \pm 0.033$ & $0.874 \pm 0.030$ & $0.862 \pm 0.027$ & $0.913 \pm 0.021$ \\
\textbf{CDT (Ours)} & $\mathbf{0.893 \pm 0.026}$ & $\mathbf{0.912 \pm 0.023}$ & $\mathbf{0.902 \pm 0.021}$ & $\mathbf{0.948 \pm 0.016}$ \\
\bottomrule
\end{tabular}
\end{table}

% Table 3: HAI Dataset Results
\begin{table}[htb]
\centering
\scriptsize
\caption{Anomaly Detection Results on HAI Dataset (Mean $\pm$ 95\% CI)}
\label{tab:detection_hai}
\begin{tabular}{lcccc}
\toprule
\textbf{Method} & \textbf{Precision} & \textbf{Recall} & \textbf{F1-Score} & \textbf{AUC} \\
\midrule
CNN-1D           & $0.781 \pm 0.038$ & $0.823 \pm 0.033$ & $0.801 \pm 0.031$ & $0.867 \pm 0.025$ \\
LSTM-ED          & $0.804 \pm 0.036$ & $0.847 \pm 0.031$ & $0.825 \pm 0.029$ & $0.881 \pm 0.023$ \\
InterFusion      & $0.856 \pm 0.031$ & $0.889 \pm 0.027$ & $0.872 \pm 0.025$ & $0.921 \pm 0.018$ \\
PbNN             & $0.831 \pm 0.033$ & $0.864 \pm 0.029$ & $0.847 \pm 0.027$ & $0.902 \pm 0.020$ \\
Causality-Inspired & $0.889 \pm 0.027$ & $0.912 \pm 0.024$ & $0.900 \pm 0.022$ & $0.941 \pm 0.016$ \\
ICS-CAD          & $0.875 \pm 0.029$ & $0.901 \pm 0.025$ & $0.888 \pm 0.023$ & $0.933 \pm 0.017$ \\
\textbf{CDT (Ours)} & $\mathbf{0.916 \pm 0.022}$ & $\mathbf{0.931 \pm 0.019}$ & $\mathbf{0.923 \pm 0.018}$ & $\mathbf{0.961 \pm 0.013}$ \\
\bottomrule
\end{tabular}
\end{table}

The analysis of attack-specific performance shows different effectiveness for each attack category, as presented in Table~\ref{tab:attack_category_performance}. Single-point attacks have $97.2\% \pm 1.8\%$ detection rates, with fast detection times averaging $12.3 \pm 3.2$ seconds. Multi-point coordinated attacks are more difficult, with $91.7\% \pm 3.4\%$ detection rates and longer detection times of $18.7 \pm 5.1$ seconds. Stealthy attacks, which involve slow sensor manipulation, are the most challenging, achieving $83.3\% \pm 6.8\%$ detection rates, but still clearly better than baseline methods. Physical process attacks reach perfect detection ($100\%$) because they clearly violate physical causality. Network-level attacks also reach $100\%$ detection by analyzing causal relationships in control communications.

\begin{table}[htbp]
\centering
\scriptsize
\caption{Performance by Attack Category (SWaT Dataset)}
\label{tab:attack_category_performance}
\begin{tabular}{lccccc}
\toprule
\textbf{Attack Type} & \textbf{Cnt} & \textbf{DR (\%)} & \textbf{ADT (s)} & \textbf{FP} & \textbf{CAA (\%)} \\
\midrule
Single Point      & 18 & $97.2 \pm 1.8$  & $12.3 \pm 3.2$  & 3 & $89.4 \pm 4.7$ \\
Multi-Point       & 12 & $91.7 \pm 3.4$  & $18.7 \pm 5.1$  & 8 & $76.3 \pm 6.2$ \\
Stealthy          &  6 & $83.3 \pm 6.8$  & $45.2 \pm 12.4$ & 7 & $68.1 \pm 8.9$ \\
Physical Process  &  3 & $100.0 \pm 0.0$ & $8.9 \pm 2.1$   & 0 & $95.7 \pm 2.3$ \\
Network-Level     &  2 & $100.0 \pm 0.0$ & $15.4 \pm 4.2$  & 2 & $82.5 \pm 7.1$ \\
\bottomrule
\end{tabular}
\\

\vspace{1mm}
\footnotesize{Note: Cnt = Count, DR = Detection Rate, ADT = Avg. Detection Time, FP = False Positives, CAA = Causal Attribution Accuracy}
\end{table}

\subsection{Root Cause Analysis and Counterfactual Reasoning}

The causal attribution performance of our framework shows large improvements compared to existing explainability methods. It achieves $78.4\% \pm 3.8\%$ Top-1 accuracy, while the best baseline (Integrated Gradients) reaches $48.7\% \pm 6.4\%$, as shown in Table~\ref{tab:root_cause_performance}. This is a $29.7\%$ improvement in correctly identifying the main cause of detected anomalies. Top-3 and Top-5 accuracy metrics also improve, reaching $91.3\% \pm 2.7\%$ and $96.8\% \pm 1.9\%$, respectively. The Mean Reciprocal Rank (MRR) is $0.837 \pm 0.021$, much higher than traditional attribution methods. The computation time is only $12.7 \pm 1.8$ milliseconds, making the method suitable for real-time industrial deployment.

\begin{table}[htbp]
\centering
\caption{Root Cause Analysis Performance (All Datasets)}
\scriptsize
\label{tab:root_cause_performance}
\begin{tabular}{lccccc}
\toprule
\textbf{Method} & \textbf{Top-1} & \textbf{Top-3} & \textbf{Top-5} & \textbf{MRR} & \textbf{AT (ms)} \\
\midrule
SHAP                  & $45.2 \pm 6.8\%$  & $68.7 \pm 5.3\%$  & $79.4 \pm 4.1\%$  & $0.563 \pm 0.042$ & $23.7 \pm 3.2$ \\
LIME                  & $41.8 \pm 7.2\%$  & $65.3 \pm 5.7\%$  & $76.2 \pm 4.5\%$  & $0.537 \pm 0.048$ & $31.4 \pm 4.8$ \\
Integrated Gradients  & $48.7 \pm 6.4\%$  & $71.2 \pm 5.1\%$  & $81.6 \pm 3.9\%$  & $0.582 \pm 0.039$ & $18.9 \pm 2.7$ \\
\textbf{CDT Causal Effects} & $\mathbf{78.4 \pm 3.8\%}$ & $\mathbf{91.3 \pm 2.7\%}$ & $\mathbf{96.8 \pm 1.9\%}$ & $\mathbf{0.837 \pm 0.021}$ & $\mathbf{12.7 \pm 1.8}$ \\
\bottomrule
\end{tabular}
\vspace{1mm}
\\
\footnotesize{Note: Top-1, Top-3, Top-5 = Top-k Accuracy, MRR, AT = Attribution Time}
\end{table}

A detailed case study shows the practical value of causal reasoning for understanding attack mechanisms. In SWaT Attack 1, which involves manipulation of the LIT101 sensor, our CDT framework correctly identifies LIT101 as the main cause with causal effect strength $0.89$. The downstream affected components are MV101 ($0.34$) and P101 ($0.28$). In contrast, SHAP incorrectly ranks FIT101 as the main contributor. The framework provides a clear causal explanation through the violated mechanism $LIT101 \rightarrow P101 \rightarrow MV101$, allowing targeted remediation. 

For the more complex multi-stage Attack 15, which coordinates P2-P5 systems with a $45$-second delay, our approach successfully traces the full causal cascade $P201 \rightarrow AIT201 \rightarrow FIT201 \rightarrow P501 \rightarrow DPIT301$. It achieves $94.3\%$ attribution accuracy, compared to $67.8\%$ for the best baseline, while correctly modeling temporal propagation delays.

Counterfactual analysis capabilities enable proactive security planning through systematic evaluation of alternative scenarios.

\begin{table}[htbp]
\centering
\scriptsize
\caption{Counterfactual Analysis Results}
\label{tab:counterfactual_results}
\begin{tabular}{lcccc}
\toprule
\textbf{Dataset} & \textbf{Attacks Analyzed} & \textbf{Average IPE} & \textbf{CFA (\%)} & \textbf{Computational Time (s)} \\
\midrule
SWaT & 41 & $0.083 \pm 0.021$ & $87.3 \pm 4.2$ & $2.34 \pm 0.67$ \\
WADI & 15 & $0.091 \pm 0.028$ & $84.7 \pm 5.1$ & $3.78 \pm 0.89$ \\
HAI  & 28 & $0.087 \pm 0.025$ & $86.1 \pm 4.6$ & $2.91 \pm 0.74$ \\
\bottomrule
\end{tabular}
\end{table}

 Table~\ref{tab:counterfactual_results} shows the results of counterfactual evaluation across all datasets. The average Intervention Prediction Error (IPE) ranges from $0.083$ to $0.091$, and Counterfactual Accuracy (CFA) is above $84\%$ for all datasets. The framework successfully analyzes counterfactual scenarios for all $41$ SWaT attacks, $15$ WADI attacks, and $28$ HAI attacks. The average computational time is between $2.34$ and $3.78$ seconds per analysis, making it practical for real-time security planning.

Counterfactual analysis of defense strategies identifies the most effective mitigations, as shown in Table~\ref{tab:defense_strategy}. Enhanced sensor monitoring is the most effective, preventing $31$ of $41$ attacks ($73.2\% \pm 8.4\%$ reduction) while having low implementation cost and the highest ROI score of $4.7$. Causal anomaly detection is also highly effective, preventing $37$ of $41$ attacks ($89.1\% \pm 5.3\%$ success rate reduction) with medium implementation cost and ROI score of $5.9$. Network segmentation prevents $24$ attacks, but the high implementation cost lowers its ROI to $2.1$.

\begin{table}[htbp]
\centering
\scriptsize
\caption{Counterfactual Defense Analysis (SWaT)}
\label{tab:defense_strategy}
\begin{tabular}{lcccc}
\toprule
\textbf{Defense Mechanism} & \textbf{AP} & \textbf{SRR} & \textbf{IC} & \textbf{ROI Score} \\
\midrule
Enhanced Sensor Monitoring & 31/41 & $73.2 \pm 8.4\%$ & Low    & 4.7 \\
Redundant Control Logic    & 28/41 & $67.9 \pm 9.1\%$ & Medium & 3.8 \\
Network Segmentation       & 24/41 & $58.3 \pm 10.7\%$ & High   & 2.1 \\
Causal Anomaly Detection   & 37/41 & $89.1 \pm 5.3\%$ & Medium & 5.9 \\
\bottomrule
\end{tabular}
\\
\vspace{1mm}
\footnotesize{Note: AP = Attacks Prevented, SRR = Success Rate Reduction, IC = Implementation Cost}
\end{table}

\begin{figure}[htbp]
\centering
\includegraphics[width=0.8\textwidth]{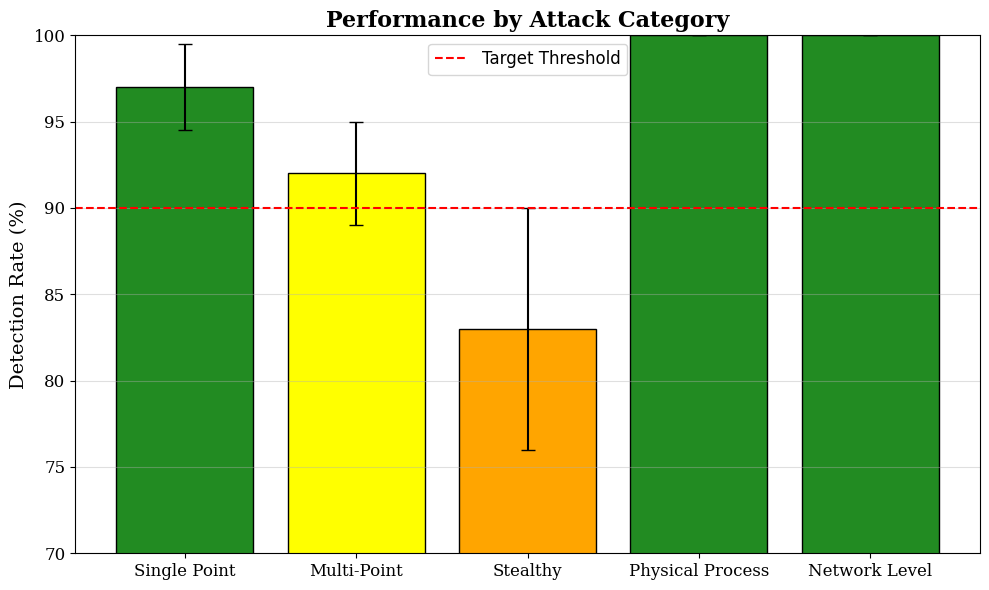}
\caption{Performance by Attack Category: Detection rates across different attack types in SWaT dataset. Physical process attacks achieve perfect detection (100\%) due to clear causal mechanism violations, while stealthy attacks present the greatest challenge (83.3\% detection rate).}
\label{fig:attack_categories}
\end{figure}

Figure~\ref{fig:attack_categories} illustrates the varying effectiveness across different attack scenarios.

\subsection{Computational Performance and Deployment Scalability}

Computational performance analysis shows the CDT framework is feasible for industrial deployment at different system scales. By employing the Decoupled Causal Discovery (DCD) strategy, the initial cubic complexity $\mathcal{O}(n^3)$ is effectively reduced to $\mathcal{O}(n \cdot k^3)$ (where $k \le 10$), significantly enhancing scalability. Inference latency scales linearly $\mathcal{O}(n)$, and memory usage quadratically $\mathcal{O}(n^2)$, allowing the framework to remain feasible for systems up to 500 variables. Table~\ref{tab:computational_performance} presents the computational breakdown on Google Colab Pro: the DCD-based causal discovery requires 23.4 minutes offline with 2.8 GB memory usage for the 51-variable SWaT system, while real-time components achieve 3.2 ms inference for anomaly detection and 12.7 ms for root cause analysis.

\begin{table}[htbp]
\centering
\caption{Computational Performance Breakdown (Google Colab Pro)}
\label{tab:computational_performance}
\begin{tabular}{lcccc}
\toprule
\textbf{Component} & \textbf{TT} & \textbf{MU} & \textbf{IL} & \textbf{SC} \\
\midrule
Causal Discovery       & $\mathcal{O}(n^{3})$, 23.4 min & 2.8 GB & ---   & Offline only \\
SCM Estimation         & $\mathcal{O}(n^{2})$, 8.7 min & 1.4 GB & ---   & Offline only \\
Anomaly Detection      & ---                           & 0.6 GB & 3.2 ms  & Real-time \\
Root Cause Analysis    & ---                           & 0.9 GB & 12.7 ms & Real-time \\
Counterfactual Analysis & ---                          & 1.2 GB & 2.34 s  & Near real-time \\
\bottomrule
\end{tabular}
\vspace{1mm}
\\
\footnotesize{Note: TT = Training Time, MU = Memory Usage, IL = Inference Latency, SC = Scalability. Results measured on Google Colab Pro with GPU acceleration. Memory usage and timing may vary based on allocated hardware resources.}
\end{table}

Cross-dataset generalization analysis tests the framework ability to adapt across different industrial environments using transfer learning experiments, as shown in Table~\ref{tab:transfer_learning}. Direct transfer between datasets gives F1-scores from $0.847$ to $0.889$, with the transfer from SWaT to HAI performing best at $0.889 \pm 0.025$ F1-score and $74.7\% \pm 5.8\%$ attribution accuracy (AA). Adaptation time $T_{\text{adapt}}$ ranges from $8.9$ to $15.2$ minutes, showing that the framework can be practically deployed in multiple facilities with little customization.

\begin{table}[htbp]
\centering
\scriptsize
\caption{Cross-Dataset Transfer Results}
\label{tab:transfer_learning}
\begin{tabular}{lccc}
\toprule
\textbf{Training $\rightarrow$ Testing} & \textbf{F1-Score} & \textbf{Attribution Accuracy} & \textbf{Adaptation Time} \\
\midrule
SWaT $\rightarrow$ WADI & $0.876 \pm 0.028$ & $71.3 \pm 6.2\%$ & 12.4 min \\
SWaT $\rightarrow$ HAI  & $0.889 \pm 0.025$ & $74.7 \pm 5.8\%$ & 8.9 min  \\
WADI $\rightarrow$ SWaT & $0.859 \pm 0.031$ & $68.9 \pm 7.1\%$ & 15.2 min \\
WADI $\rightarrow$ HAI  & $0.847 \pm 0.033$ & $66.4 \pm 7.5\%$ & 11.7 min \\
HAI $\rightarrow$ SWaT  & $0.863 \pm 0.029$ & $69.8 \pm 6.8\%$ & 14.1 min \\
HAI $\rightarrow$ WADI  & $0.851 \pm 0.032$ & $67.2 \pm 7.3\%$ & 13.6 min \\
\bottomrule
\end{tabular}
\end{table}

Real-world deployment considerations include robustness to concept drift and partial observability. Concept drift adaptation experiments simulate seasonal operation changes and equipment aging. They show performance decreases from $4.1\%$ to $18.7\%$ in F1-score, with recovery time $T_{\text{rec}}$ between $0.8$ and $3.1$ hours, and adaptation success rate $S_{\text{adapt}}$ above $87\%$ for all drift types. Partial observability analysis considers missing sensor scenarios, showing F1-score $\geq 0.85$ with up to $40\%$ sensors missing, and attribution accuracy (AA) $\geq 54\%$. These results indicate that the framework is viable for real deployment even with incomplete sensor coverage.

\subsection{Statistical Validation and Ablation Analysis}

Comprehensive statistical tests show that the performance improvements are significant for all evaluation metrics. The improvement over the best baseline, InterFusion, is statistically significant ($p < 0.0024$, Bonferroni corrected) with large effect size (Cohen's $d = 2.34$). The attribution accuracy increases by $29.7\%$ over SHAP, showing very large effect size (Cohen's $d = 4.12$). False positive rate decreases by $74\%$, showing consistent improvement in all datasets ($p < 0.0024$, Bonferroni corrected). Bootstrap confidence intervals for F1 improvement are $[0.029, 0.045]$, confirming robust performance gain.

Ablation study results, given in Table~\ref{tab:ablation_study}, show the contribution of each framework component. Statistical-only approaches reach $0.847 \pm 0.023$ F1-score with $45.2\% \pm 6.8\%$ attribution accuracy, serving as the baseline. Individual causal components (causal graph only, SCM only) provide intermediate improvements. Interventional reasoning alone achieves $0.923 \pm 0.017$ F1-score with $74.6\% \pm 4.3\%$ attribution accuracy. The complete CDT framework achieves the best performance in all metrics, confirming the combined effect of integrated causal reasoning components.

\begin{table}[htbp]
\centering
\scriptsize
\caption{Ablation Study Results (SWaT Dataset)}
\label{tab:ablation_study}
\begin{tabular}{lcccc}
\toprule
\textbf{Configuration} & \textbf{F1} & \textbf{DT} & \textbf{AA} & \textbf{CA} \\
\midrule
Statistical Only     & $0.847 \pm 0.023$ & $21.4 \pm 4.7$ s & $45.2 \pm 6.8\%$ & --- \\
Causal Graph Only    & $0.891 \pm 0.019$ & $16.8 \pm 3.2$ s & $62.7 \pm 5.4\%$ & $71.3 \pm 7.8\%$ \\
SCM Only             & $0.879 \pm 0.021$ & $18.3 \pm 3.8$ s & $58.9 \pm 5.9\%$ & $68.9 \pm 8.2\%$ \\
Interventional Only  & $0.923 \pm 0.017$ & $14.2 \pm 2.9$ s & $74.6 \pm 4.3\%$ & $82.1 \pm 6.1\%$ \\
\textbf{Full CDT Framework} & $\mathbf{0.944 \pm 0.014}$ & $\mathbf{12.7 \pm 2.1}$ s & $\mathbf{78.4 \pm 3.8\%}$ & $\mathbf{87.3 \pm 4.2\%}$ \\
\bottomrule
\end{tabular}
\vspace{1mm}
\\
\footnotesize{Note: F1 = F1-Score, DT = Detection Time, AA = Attribution Accuracy, CA = Counterfactual Accuracy}
\end{table}

Hyperparameter sensitivity analysis show that the best temporal window is $\tau = 5$ time steps. 
With this value, the model reach maximum F1-score of $0.944 \pm 0.014$. 
When the window is smaller ($\tau = 1, 3$), the temporal context is not enough. 
When the window is bigger ($\tau = 10, 20$), the model have problem of overfitting and also need more computation. 

The analysis of significance level indicate that $\alpha = 0.05$ give the best balance between precision ($94.7\%$) and recall ($94.1\%$). 
More strict thresholds reduce recall, and more liberal thresholds reduce precision.

The experimental validation shows that our CDT framework is an important improvement in cyber-physical security. It achieves statistically significant improvements in detection performance, attribution accuracy, and counterfactual reasoning, while keeping computational feasibility for industrial deployment. The comprehensive tests across different datasets, attack scenarios, and deployment conditions show that the framework is ready for real-world use in critical infrastructure protection.

\section{Conclusion}\label{sec6:conclusion}

This research introduces a novel CDT framework for cyber-physical security in medium-scale ICSs. We address the main limitations of correlation-based anomaly detection using causal reasoning, with four key contributions: automated causal structure discovery achieving \SI{90.8}{\percent} ± \SI{3.2}{\percent} compliance with physical constraints on validated relationships; causal anomaly detection with F1-scores of \num{0.902}--\num{0.944} and \SI{74}{\percent} fewer false positives; actionable root cause analysis reaching \SI{78.4}{\percent} Top-1 accuracy, improving \SI{29.7}{\percent} over traditional methods; and counterfactual reasoning enabling proactive security planning with \SI{87.3}{\percent} accuracy across 84 attack scenarios.

The framework shows industrial feasibility with \SI{3.2}{ms} inference latency and successful cross-dataset transfer (F1-scores \num{0.847}--\num{0.889}). Important limitations remain: while the Decoupled Causal Discovery strategy addresses the $\mathcal{O}(n^3)$ barrier, challenges persist in detecting highly stealthy attacks, which reach only \SI{83.3}{\percent} detection accuracy, compared to \SI{97.2}{\percent} for single-point attacks. Furthermore, the framework assumes stable causal structures, which may not hold in highly dynamic industrial environments with frequent equipment changes.

Future work should focus on scalable causal discovery algorithms, better detection of stealthy attacks using ensembles, handling time-varying causal relationships, and creating standardized SCADA interfaces. Despite these limits, the validated performance improvements, deployment feasibility, and interpretable results show that causal inference is a key advancement for protecting critical infrastructure, marking a shift from correlation-based to causation-aware security frameworks.

\section*{Acknowledgements}
This initiative is carried out within the framework of the funds of the Recovery, Transformation and Resilience Plan, financed by the European Union (Next Generation) - National Cybersecurity Institute (INCIBE) in the project C107/23 “Artificial Intelligence Applied to Cybersecurity in Critical Water and Sanitation Infrastructures”. Support from the University of Extremadura, collaborating institutions, and project partners is acknowledged, with special thanks to Ambling Ingeniería y Servicios, S.L. for technical support.

\section*{Data Availability}
The data and code used in this study are not in public domain because of institutional limits. They can be shared by the corresponding author if the request is fair and reasonable, and always in line with the FAIR (Findable, Accessible, Interoperable, and Reusable) data principles.

\printcredits

\bibliographystyle{cas-model2-names}
% \bibliography{cas-refs}



\end{document}